\documentclass[conference,10pt,letter]{IEEEtran}
\usepackage{amsthm,amsmath,comment,bbm,amssymb,color,graphicx}
\usepackage{mathtools}
\usepackage{comment}
\usepackage{url}
\usepackage{caption}
\usepackage{subcaption}
\usepackage{epstopdf}
\usepackage{longtable}
\usepackage{cite}
\usepackage{tikz}
\usetikzlibrary{shapes}
\usepackage{multirow}
\usepackage{multicol}
\usepackage[top=0.7in,left=0.673in,right=0.673in,bottom=1.02in]{geometry}

\def\nbh{{\mathbf{h}}}

\def\nbp{{\mathbf{p}}}

\def\nbu{{\mathbf{u}}}

\def\nb0{{\mathbf{0}}}
\def\nb1{{\mathbf{1}}}

\def\ncalI{{\mathcal{I}}}

\def\ncalM{{\mathcal{M}}}

\def\nbbM{{\mathbb{M}}}

\def\nbbP{{\mathbb{P}}}

\newtheorem{lem}{Lemma}

\newtheorem{nrem}{Remark}

\newtheorem{cor}{Corollary}

\newtheorem{eg}{Example}

\IEEEoverridecommandlockouts
\begin{document}
\title{Rate-Splitting Multiple Access for Non-Orthogonal Unicast Multicast: An Experimental Study}

\author{\IEEEauthorblockN{Xinze Lyu, Sundar Aditya and Bruno Clerckx}\\[-2.0ex]
\IEEEauthorblockA{Dept. of Electrical and Electronic Engineering, Imperial College London, London SW7 2AZ, UK}\\[-2.0ex]
\IEEEauthorblockA{Email:\{x.lyu21, s.aditya, b.clerckx\}@imperial.ac.uk}\\[-6.0ex]
}
\maketitle

\begin{abstract}
Non-orthogonal unicast multicast (NOUM) is a variant of multi-antenna multi-user communications where the users desire a shared message (multicast) in addition to their respective unique messages (unicast). The multicast rate is capped in many emerging NOUM applications, such as live-event broadcasting, location-based services and vehicular communications. Given this constraint, we experimentally show that when the user channels are highly correlated, Rate-Splitting Multiple Access (RSMA)-based NOUM can meet the multicast rate while supporting a larger unicast sum rate than multi-user linear precoding (MULP)-based NOUM and orthogonal unicast multicast.
\end{abstract}


\bstctlcite{IEEEexample:BSTcontrol}

\section{Introduction}
Non-orthogonal Unicast Multicast (NOUM) is a unique form of multi-antenna multi-user communications where users desire both multicast (shared) and unicast (unique) messages. These could arise in live-event broadcasting (where the unicast messages could be social media interactions), Vehicle-to-Everything (V2X) networks and Augmented Reality/Virtual Reality (AR/VR) applications (where the multicast message could be location-dependent information), etc. NOUM is a spectrally efficient way to jointly realize (physical layer) multicast and unicast in multi-antenna systems, with great potential for next-general wireless networks  \cite{Erricsson_broadcast_multicast}, provided adequate multicast and unicast data rates can be guaranteed.

The conventional way of realizing NOUM is by extending Multi-User Linear Precoding (MULP) \cite{Sethakaset_pimrc_2010, Liu_spawc_2017, Wang_mmwave_noum_2018, Zhao_TBC_2020}; i.e., apart from one precoder per unicast message, there is also a multicast precoder. The precoder design amounts to superposition coding at the transmitter (TX), with each user decoding two messages (the multicast message followed by its unicast message) via successive interference cancellation (SIC).

The above scheme closely resembles Rate-Splitting Multiple Access (RSMA), which has been shown to improve the performance of multi-antenna multi-user communications in several aspects, such as spectral efficiency, energy efficiency, mobility etc. \cite{RSMA_JSAC_Primer, mao2022fundmental}.  In particular, \emph{unicast} RSMA involves the creation of $K+1$ data streams and precoders for $K$ users. The additional stream -- known as the \emph{common stream} -- is composed of parts of each user's message, as illustrated in Table~\ref{tab:rsma_noum_subtleties}.  Similar to MULP-based NOUM, each user employs SIC to first decode the common stream ($s_c$), followed by its so-called \emph{private stream} ($s_i, i\in\{1,2\}$) to retrieve its desired message. This two-step retrieval process allows each user to partially decode and subtract the inter-user interference (through the common stream), and partially treat it as noise (i.e., the private streams of other users). This, in turn, increases the sum rate \cite{HamdiMISOImperfectCSIT, lyu2023prototype}.

In fact, MULP-based NOUM can be viewed as a special case of RSMA, where the common stream carries only the multicast message instead of a mixture of unicast messages. However, RSMA offers more flexibility for NOUM w.r.t the composition of the common stream, which translates to better rate \cite{Lina_NOUM_TCOM} and latency \cite{RSMA_v2x_federated} performance. To see how, two points need to be noted:
\begin{itemize}
    \item[(a)] In many NOUM applications, the multicast rate is capped (e.g., video streaming, V2X safety messages, etc. \cite{3GPP_5g_evo_multi_broad}). Hence, a well-motivated objective is to maximize the unicast sum rate, subject to satisfying the multicast rate.

    \item[(b)] The common stream in RSMA unicast allows all users to partially decode and eliminate the interference from other users.
\end{itemize}
In light of (a), MULP-based NOUM only allocates enough resources (e.g., precoder power) to the common stream to meet the desired multicast rate. However, with RSMA-based NOUM, the common stream can be resourced to support a higher rate than the multicast rate, with the difference contributing to the unicast sum rate. This can be realized by designing the common stream to be a mixture of the multicast message and parts of each user's unicast message, as shown in Table~\ref{tab:rsma_noum_subtleties}. Just like (b), the unicast message components in the common stream help with interference suppression, which can increase the unicast sum rate beyond what is achievable with MULP-based NOUM.

While the benefits of RSMA-based NOUM have been investigated analytically in \cite{Lina_NOUM_TCOM, RSMA_v2x_federated}, an experimental validation of the same is absent in the literature. This motivates our efforts in this paper, where we first characterize the conditions under which RSMA-based NOUM achieves a higher unicast sum rate than MULP-based NOUM while satisfying a given multicast rate.  We then validate this experimentally and find that the gains of RSMA-based NOUM are more pronounced with increasing channel correlation among the users.

\begin{table}[]
\setlength{\tabcolsep}{4pt} 
\renewcommand{\arraystretch}{1} 
    \centering
    \begin{tabular}{|c|c|c|c|}
    \hline
    Stream & MULP-based NOUM & RSMA unicast &RSMA-based NOUM \\
    \hline
     $s_c$ & $W_0$ & $[W_{c,1}, W_{c,2}]$ & $[W_0, W_{c,1}, W_{c,2}]$ \\
     $s_1$ & $W_1$ & $W_{p,1}$ & $W_{p,1}$ \\
     $s_2$ & $W_2$ & $W_{p,2}$ & $W_{p,2}$ \\
     \hline
    \end{tabular}
    \caption{Comparison between MULP-based NOUM, RSMA unicast and RSMA-based NOUM for the two-user case. $W_0$ is the multicast message, and $W_i$ is the unicast message for user $i~(= 1,2)$. For RSMA, $W_i:= [W_{c,i}, W_{p,i}]$. MULP-based NOUM is a special (and in general, suboptimal) case of RSMA-based NOUM.}
    \label{tab:rsma_noum_subtleties}
\end{table}

\section{System Model for RSMA-based NOUM}
We follow the RSMA-based NOUM architecture of \cite{Lina_NOUM_TCOM}. Consider the two-user multiple-input single-output (MISO) case. Let $W_0$ denote the multicast message, and $W_i~(i=1,2)$ the unicast message for user $i$. At the TX, each $W_i$ is split into common and private portions, denoted by $W_{c,i}$ and $W_{p,i}$, respectively (in general, the splitting ratio need not be 50-50 nor does it have to be the same across both users). Let $W_c := [W_0, W_{c,1}, W_{c,2}]$ denote the common message formed by concatenating the multicast message along with the common portions of each unicast message. Through an appropriately chosen Modulation and Coding Scheme (MCS) level, $W_c$ is transformed into a common stream, $s_c$. Similarly, $W_{p,1}$ and $W_{p,2}$ are separately encoded and modulated to form private streams, $s_1$ and $s_2$, respectively. All three streams are linearly precoded to form the transmit signal:
    \begin{equation}
        \label{eq:transmit stream equation}
        \mathbf{x} = \mathbf{p}_c s_c + \mathbf{p}_1 s_1 + \mathbf{p}_2 s_2,
    \end{equation}
where $\nbp_c$ denotes the common stream precoder, and $\nbp_i$ the private stream precoder of user~$i$.

Let $\nbh_i$ denote the channel between the TX and the $i$-th user. The received signal, $y_i$, at the $i$-th user is given by:
    \begin{align}
        \label{eq:RX k received signal}
        y_i &= \mathbf{h}_i^H\mathbf{x} + n_i \hspace{3mm} (i=1,2)\notag \\
            &= \nbh_i^H \nbp_c s_c + \nbh_i^H \nbp_1 s_1 +  \nbh_i^H \nbp_2 s_2 + n_i ,
    \end{align}
where $n_i$ is the receiver thermal noise. Upon receiving $y_i$, the $i$-th user first decodes $s_c$, while treating the interference from the $s_1$ and $s_2$ as noise. This enables the $i$-th user to recover $W_0$ and a part of its unicast message ($W_{c,i}$). Then, through SIC, the $i$-th user decodes $s_i$ to recover the rest of its unicast message ($W_{p,i}$) by treating the interference from the other private stream as noise. It is easy to see from (\ref{eq:RX k received signal}) that the achievable multicast and unicast rates are a function of the precoders and MCS levels chosen for each of the three streams. We describe these next. 
\begin{itemize}
    \item \textbf{Precoder Design}: Let $\hat{\nbh}_i$ denote the estimate of $\nbh_i$ at the TX. 
    Since $s_c$ must be decoded by both users, $\nbp_c$ must have sufficient gain along each $\hat{\nbh}_i$. Thus, weighted maximum-ratio transmission (MRT) precoding is a reasonable good heursitic choice for $\nbp_c$, i.e.,
\begin{align}
    \label{eq:pc_mrt}
    \nbp_c(t) &:= \sqrt{P_t (1-t)} \frac{(\hat{\nbu}_1 + \hat{\nbu}_2)}{\|\hat{\nbu}_1 + \hat{\nbu}_2\|},
\end{align}
where $\hat{\nbu}_i$ denotes the unit vector along $\hat{\nbh}_i$. On the other hand, user $i$ decodes $s_i$ by treating the interference from $s_j~(j\neq i)$ as noise. Hence, to suppress this interference as much as possible, we consider zero-forcing (ZF) precoding for the private streams, i.e.,
\begin{align}
    \label{eq:pi_zf}
    \nbp_i(t) &:= \sqrt{P_t (t/2)} \hat{\nbu}_{j}^{\perp} ~(j \neq i; i = 1,2), 
\end{align}
where $\nbu_i^\perp$ denotes the unit vector along the nullspace of $\hat{\nbh}_i$. In (\ref{eq:pc_mrt}) and (\ref{eq:pi_zf}), the parameter $t\in[0,1]$ determines the allocation of the transmit power, $P_t$, to the three streams.

    \item \textbf{MCS}: An MCS level can be defined by a pair $\ncalI = (m, r)$, where positive integer $m$ denotes the bits per modulation symbol (e.g., 2 for QPSK) and $r \in (0,1]$ denotes the code rate. In general, the MCS level of each stream can be distinct. Hence, let $\ncalI_c = (m_c, r_c)$, $\ncalI_1 = (m_1, r_1)$ and $\ncalI_2 = (m_2, r_2)$ denote the MCS levels chosen for $s_c$, $s_1$ and $s_2$, respectively. 
\end{itemize}

Let $\ncalM = \{\ncalI_c, \ncalI_1, \ncalI_2\}$. As a function of $t$ and $\ncalM$, let $R_c(t, \ncalM)$ and $R_i(t, \ncalM)$ denote the \emph{MCS-limited rate} (in bits/s/Hz) for $s_c$ and $s_i$, respectively. Their expressions are as follows:
\begin{align}
\label{eq:Rc}
    R_c (t, \ncalM) &= m_c r_c \times \nbbP(\hat{W}_c^{(1)} = W_c; \hat{W}_c^{(2)} = W_c) \\
\label{eq:Ri}
    R_i (t, \ncalM) &= m_1 r_1 \times \nbbP(\hat{W}_{p,i} = W_{p,i}) ~ (i = 1,2) ,
\end{align}
where the probability terms capture the rate loss due to decoding errors. In particular, $\nbbP(\hat{W}_c^{(1)} = W_c; \hat{W}_c^{(2)} = W_c)$ in (\ref{eq:Rc}) denotes the probability that $W_c$ is correctly decoded at both users, while $\nbbP(\hat{W}_{p,i} = W_{p,i})$ in (\ref{eq:Ri}) denotes the probability that $W_{p,i}$ is correctly decoded at the $i$-th user. These probabilities are difficult to characterize in closed-form, but can be evaluated empirically (see Section~\ref{sec:meas}).

Let $R_{\rm mult}^{\rm RS}(t, \ncalM)$ denote the multicast rate and $R_{\rm uni}^{\rm RS}(t, \ncalM)$ the unicast sum rate for RSMA-based NOUM. Their expressions are related to (\ref{eq:Rc}) and (\ref{eq:Ri}) as follows:
\begin{align}
\label{eq:Rmult_RS}
    R_{\rm mult}^{\rm RS}(t, \ncalM) &= \frac{|W_0|}{|W_c|} R_c(t, \ncalM) \\
\label{eq:Runi_RS}
    R_{\rm uni}^{\rm RS}(t, \ncalM) &= \left(1 - \frac{|W_0|}{|W_c|}\right) R_c(t, \ncalM) + \sum_{i = 1}^2 R_i(t,\ncalM) 
\end{align}
where $|W|$ denotes the size of message $W$.

\begin{nrem}
    \label{rem:mulp_message}
    For the special case when $W_c = W_0$, RSMA-based NOUM reduces to MULP-based NOUM (see Table~\ref{tab:rsma_noum_subtleties}). From (\ref{eq:Rmult_RS}) and (\ref{eq:Runi_RS}), let $R_{\rm mult}^{\rm MULP}(t, \ncalM)$ and $R_{\rm uni}^{\rm MULP}(t, \ncalM)$ denote the corresponding multicast and unicast sum rate, respectively.
\end{nrem} 

In the following section, we describe how RSMA-based NOUM can achieve higher unicast sum rate than MULP-based NOUM, for fixed multicast rate.

\section{Benefits of RSMA-based NOUM}
From (\ref{eq:pc_mrt})-(\ref{eq:Ri}), the sum rate for RSMA-based NOUM, denoted by $R_{\rm sum}(t,\ncalM)$, is given by:
\begin{align}
    \label{eq:Rsum_tM}
    R_{\rm sum}(t,\ncalM) := R_c(t,\ncalM) + R_1(t,\ncalM) + R_2(t,\ncalM). 
\end{align}
For fixed $t$, let
\begin{align}
    \label{eq:Rsum_t}
    R_{\rm sum}(t) &:= \max_{\ncalM \in \nbbM} ~ R_{\rm sum}(t, \ncalM) = R_c(t) + R_1(t) + R_2(t),
\end{align}
where $\nbbM$ denotes the collection of permissible MCS levels for the three streams, typically pre-determined through standards.

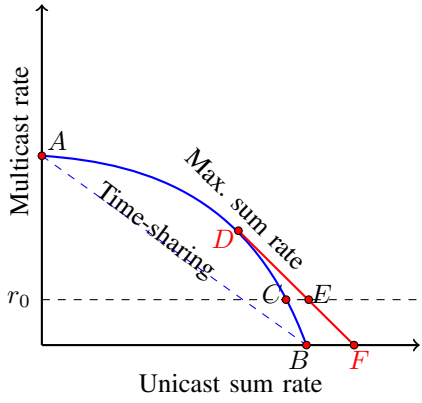
\begin{figure}
    \centering
    \begin{tikzpicture}
     \draw[thick,->] (0,0) -- (5,0) node at (2.5,-0.5){Unicast sum rate};
     \draw[thick,->] (0,0) -- (0,4.5) node[rotate = 90] at (-0.3,2.7){Multicast rate};
     \draw[dashed,blue, -] (0,2.5) -- (3.5,0) node[text=black] at (0.2,2.7){$A$}  node[text=black] at (3.4,-0.2){$B$}; 
     \draw[dashed,-] (0,0.6) -- (5,0.6) node at (-0.3,0.6){$r_0$} node at (3.05, 0.68){$C$} node at (3.7, 0.68){$E$};
     \draw[thick, blue, -] (0,2.5) .. controls (1.5,2.4) and (2.8,1.9) .. (3.5,0);
     \draw[thick,red,-] (2.6,1.51) -- (4.13,0) node at (4.2,-0.2){$F$} node at (2.4,1.4){$D$};
     \node[rotate=-37] at (1.5,1.5) {Time-sharing};
     \node[rotate=-45] at (2.7,1.8) {Max.~sum rate};
     \draw[fill=red] (0,2.5) circle (0.05);
     \draw[fill=red] (3.5,0) circle (0.05);
     \draw[fill=red] (4.13,0) circle (0.05);
     \draw[fill=red] (3.23,0.6) circle (0.05);
     \draw[fill=red] (3.53,0.6) circle (0.05);
     \draw[fill=red] (2.6,1.51) circle (0.05); 
    \end{tikzpicture}
    \caption{The achievable rate region for MULP-based NOUM (blue curve) and RSMA-based NOUM (solid red line).}
    \label{fig:region}
\end{figure}

The achievable pairs of multicast and unicast sum rates for MULP-based and RSMA-based NOUM are shown in Fig.~\ref{fig:region}. The achievable region is convex due to time-sharing. Based on Fig.~\ref{fig:region}, we make the following remarks.
\begin{nrem}[Purely Multicast]
\label{rem:ptA}
The $y$-axis points in the achievable region correspond to purely multicast, where only the common stream is transmitted. Point A represents the peak multicast rate, equal to $R_{\rm sum}(0)$ in (\ref{eq:Rsum_t}).
\end{nrem}

\begin{nrem}[Unicast SDMA]
\label{rem:ptB}
The $x$-axis from the origin till point B corresponds to unicast SDMA, with only the two private streams being transmitted. Point B represents the peak unicast SDMA sum rate, equal to $R_{\rm sum}(1)$.
\end{nrem}

\begin{nrem}[MULP-based NOUM]
\label{rem:ptC}
For $t \in (0,1)$, three streams are transmitted. Assuming perfect SIC and MCS levels of arbitrary granularity, the blue curve represents the rate boundary for MULP-based NOUM. In particular, the following multicast and unicast sum rates are achievable (point C):
\begin{align}
    R_{\rm mult}^{\rm MULP}(t) &= R_c(t) \\
    R_{\rm uni}^{\rm MULP}(t) &= R_1(t) + R_2(t)
\end{align}
\end{nrem}

\begin{nrem}[Max.~Sum Rate]
\label{rem:ptD}
Let $R_{\rm sum}^*$ denote the maximum sum rate obtained at $t^*$, corresponding to point D, i.e.,
\begin{align}
    R_{\rm sum}^* &:= R_{\rm sum}(t^*) = R_c(t^*) + R_1(t^*) + R_2(t^*).
\end{align}    
\end{nrem}

We now characterize the conditions under which RSMA-based NOUM achieves a higher unicast sum rate than MULP-based NOUM for a fixed multicast rate.
\begin{lem}
\label{lem1}
    For $t^* \in [0,1)$ and a multicast rate $r_0$ satisfying $r_0 < R_c(t^*)$, RSMA-based NOUM achieves a higher unicast sum rate than MULP-based NOUM. 
\end{lem}
\begin{IEEEproof}
From Remark~\ref{rem:ptC}, there exists $t_0$ that achieves the multicast rate, $r_0$, through MULP-based NOUM, with the resulting unicast sum rate equal to $R_1(t_0) + R_2(t_0)$, i.e.:
\begin{align}
\label{eq:RS_subopt1}
    R_{\rm mult}^{\rm MULP}(t_0) &= R_c(t_0) = r_0 \\
\label{eq:RS_subopt2}
    R_{\rm uni}^{\rm MULP}(t_0) &= R_1(t_0) + R_2(t_0)
\end{align}
Alternately, by choosing $t^*$ and allocating -- through (\ref{eq:Rmult_RS}) -- only as much of the common stream as needed to satisfy the multicast rate, RSMA-based NOUM achieves the following rates: 
\begin{align}
    R_{\rm sum}(t^*) &= R_c(t^*) + R_1(t^*) + R_2(t^*) \notag \\
                      &= R_{\rm mult}^{\rm RS}(t^*) + R_{\rm uni}^{\rm RS}(t^*)  \\
\label{eq:RS_opt1}
\mbox{where},~  R_{\rm mult}^{\rm RS}(t^*) &= r_0 \\
\label{eq:RS_opt2}
    R_{\rm uni}^{\rm RS}(t^*) &= \underbrace{R_c(t^*) - r_0}_{\geq 0} + R_1(t^*) + R_2(t^*)
 \end{align} 
The rest of the common stream is allocated to the unicast messages. This scheme corresponds to 
point \emph{E} in Fig.~\ref{fig:region}. Hence, we have
\begin{align}
    R_{\rm sum}(t_0) &\leq R_{\rm sum}(t^*)   \hspace{25mm} (\mbox{Remark~\ref{rem:ptD}}) \notag \\
    r_0 + R_{\rm uni}^{\rm MULP}(t_0) &\leq r_0 + R_{\rm uni}^{\rm RS}(t^*)  \hspace{13mm}(\mbox{from (\ref{eq:RS_subopt1})-(\ref{eq:RS_opt2})}) \notag \\
    \label{eq:final_res}
    \implies ~ R_{\rm uni}^{\rm MULP}(t_0) &\leq R_{\rm uni}^{\rm RS}(t^*). 
\end{align}

\end{IEEEproof}

\begin{cor}
\label{cor1}
For $r_0 = 0$ (i.e., purely unicast), we obtain the well-known result that the RSMA unicast sum rate is at least as large as the SDMA unicast sum rate, as captured by point F on the $x$-axis in Fig.~\ref{fig:region}.
\end{cor}

\begin{cor}
\label{cor2}
 When $t^* = 1$, RSMA-based NOUM switches to unicast only (no common stream), and RSMA-based NOUM coincides with MULP-based NOUM. The RSMA-based NOUM region (red line) and MULP-based NOUM (blue line) coincide.
\end{cor}
In the following section, we experimentally validate Lemma \ref{lem1}.

\section{Measurement Results}
\label{sec:meas}
        \begin{table}
        \centering
        \begin{tabular}{|c|c|c|c|}
        \hline
        MCS   & Modn.  & Code Rate  & Max.~Data Rate\\
        Index ($\mathcal{I}$) & ($m$)  & ($r$)     &  ($Bmr$, in Mbps) \\
        \hline
        $0, 1$         & BPSK (1)    & $1/2, 3/4$         & $6, 9$               \\ \hline
        $2, 3$         & QPSK (2)    & $1/2, 3/4$         & $12, 18$              \\ \hline
        $4, 5$         & 16QAM (4)   & $1/2, 3/4$         & $24, 36$              \\ \hline
        $6, 7$         & 64QAM (6)   & $2/3, 3/4$         & $48, 54$              \\ \hline
        $8, 9$         & 256QAM (8)  & $3/4, 5/6$         & $72, 80$              \\ \hline
        \end{tabular}
        \caption{MCS levels (largely based on IEEE 802.11g) used in our experiments. The bandwidth, $B$, equals $12{\rm MHz}$.}
        \label{tab: Mcs table}
        \end{table}
        
\begin{figure}
   \centering
   \includegraphics[width = 0.75\linewidth]{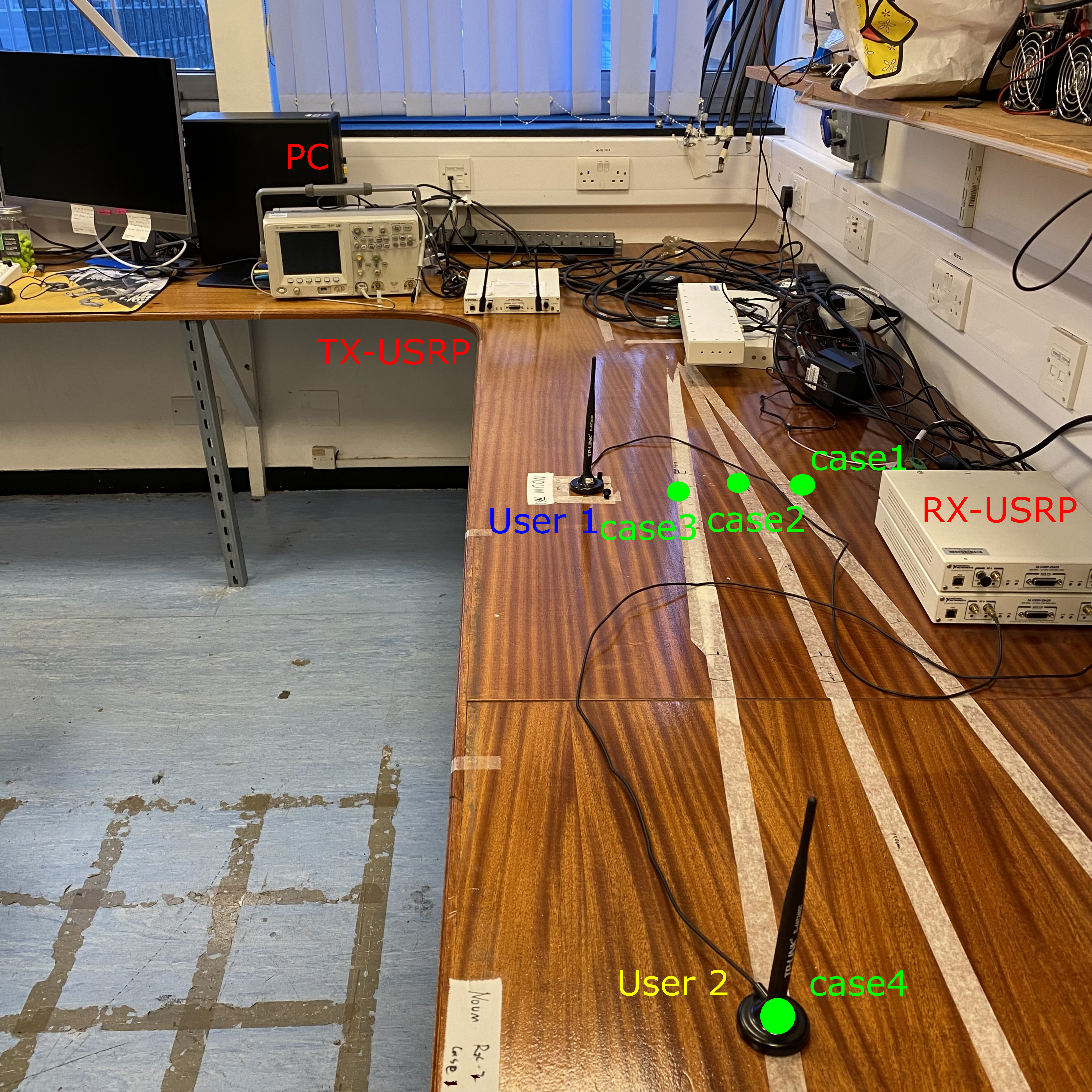}
   \caption{Layout of measurement cases. The location of the user antennas corresponds to Case 4.}
   \label{fig:NOUM_measured_cases}
\end{figure}

We realize a $2\times 2$ MISO scenario for RSMA-based NOUM using a pair of NI USRP-2942 software-defined radios. We use OFDM signals for the three streams, with the MCS levels chosen from Table~\ref{tab: Mcs table}. For details of our implementation (e.g., channel estimation, OFDM frame structure, etc.), see \cite{lyu2023prototype}. We consider four cases as shown in Fig.~\ref{fig:NOUM_measured_cases}, and described below:
\begin{itemize}
    \item \textbf{Cases 1-3}: For the first three cases, the two users experience similar pathloss, but their channel correlation progressively increases from Case 1 (least) to Case 3 (highest), as the angular separation of the users w.r.t TX narrows.
    \item \textbf{Case 4}: To study the impact of different pathloss among users, we consider an additional case where user 2 is placed 2m further away from the TX than user 1. The channel correlation is similar to Case 3.
\end{itemize}For each case and fixed $(t, \ncalM)$, we conduct 50 measurement runs, where each run involves transmitting a payload of 50 OFDM symbols per stream. 

The measured throughputs for the common stream and the $i$-th user's private stream equals $BR_c(t,\ncalM)$ and $BR_i(t,\ncalM)$, where $B$ denotes the bandwidth (equal to $12{\rm MHz}$). The probabilities in (\ref{eq:Rc}) and (\ref{eq:Ri}) are evaluated empirically (i.e., the fraction of measurement runs in which the common/private stream is successfully decoded by both/$i$-th user). The maximization in (\ref{eq:Rsum_t}) is done via brute-force search over the MCS levels in Table~\ref{tab: Mcs table}, and $t$ varies from $0$ to $1$ in steps of $0.1$. A list of parameters used for our measurements is provided in Table~\ref{tab:param_list}.
\begin{table}[]
    \centering
    \begin{tabular}{|c|c|}
    \hline 
      Parameter & Value\\
    \hline
       Center frequency  & $2.484{\rm GHz}$\\
       Transmit power, $P_t$ & $23{\rm dBm}$\\
       TX antenna length & $0.13{\rm m}$ \\
       Fraunhofer distance & $0.28{\rm m}$\\
    \hline 
       Total/Data/Pilot/Guard band subcarriers & $64/48/4/12$ \\
       Cyclic prefix length & $16$ \\
       Bandwidth, $B$ (excl. overheads) & $12{\rm MHz}$ \\ 
       OFDM symbols in payload & $50$ \\
       Experiment runs (per case) & $50$\\
       \hline       
       Distances: User 1 $\rightarrow$ TX  & $1.25{\rm m}$\\
          User 2 $\rightarrow$ TX (Case 1-3, 4)& ($1.25{\rm m}, 3.30{\rm m})$\\
    \hline
       Channel coding & Polar \\
    \hline
    \end{tabular}
    \caption{List of parameters used in our experiments.}
    \label{tab:param_list}
\end{table}
\begin{figure*}[ht]
    \begin{subfigure}{0.24\textwidth}
        \includegraphics[width=\linewidth]{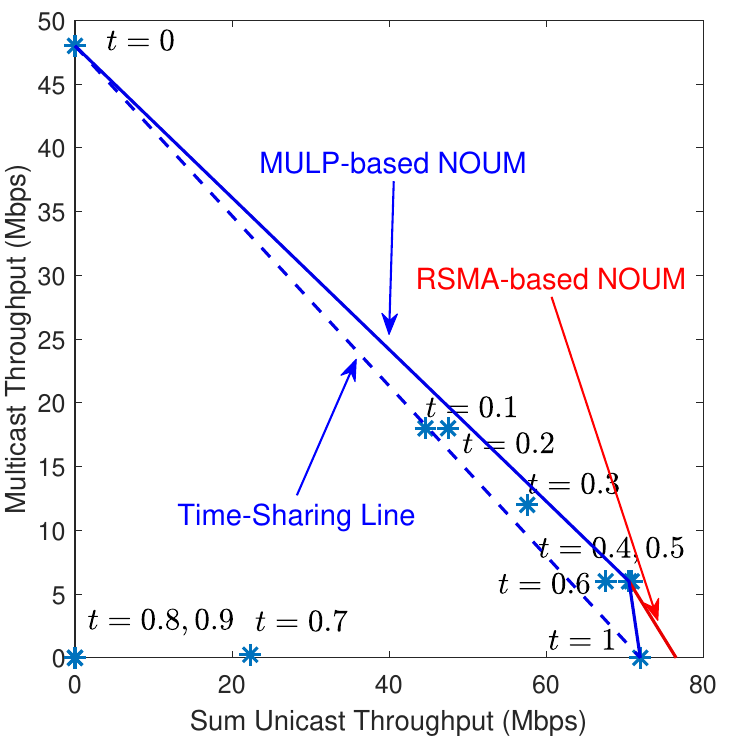}
        \caption{Case 1 ($t^*=0.5$)}
    \end{subfigure}
    \begin{subfigure}{0.24\textwidth}
        \includegraphics[width=\linewidth]{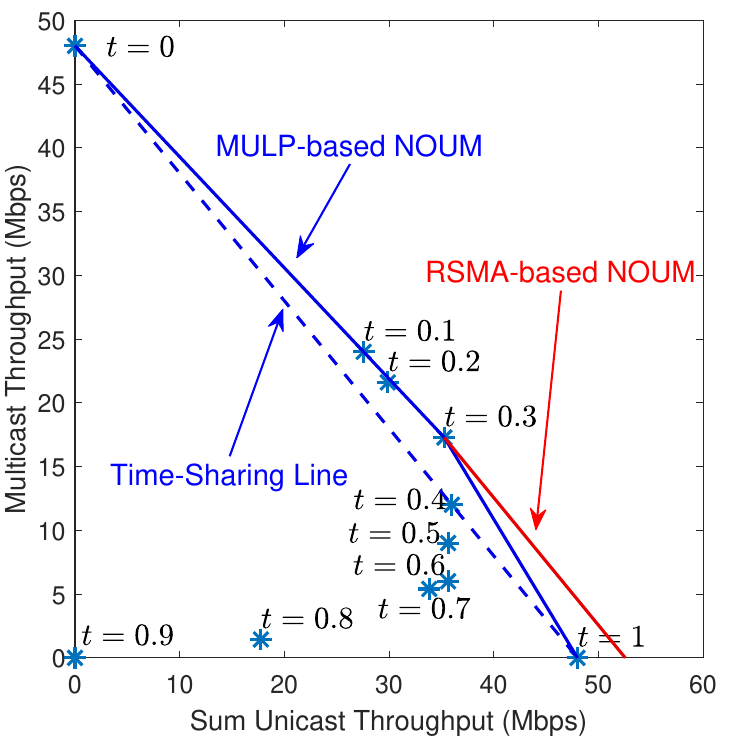}
        \caption{Case 2 ($t^*=0.3$)}
    \end{subfigure}
    \begin{subfigure}{0.24\textwidth}
        \includegraphics[width=\linewidth]{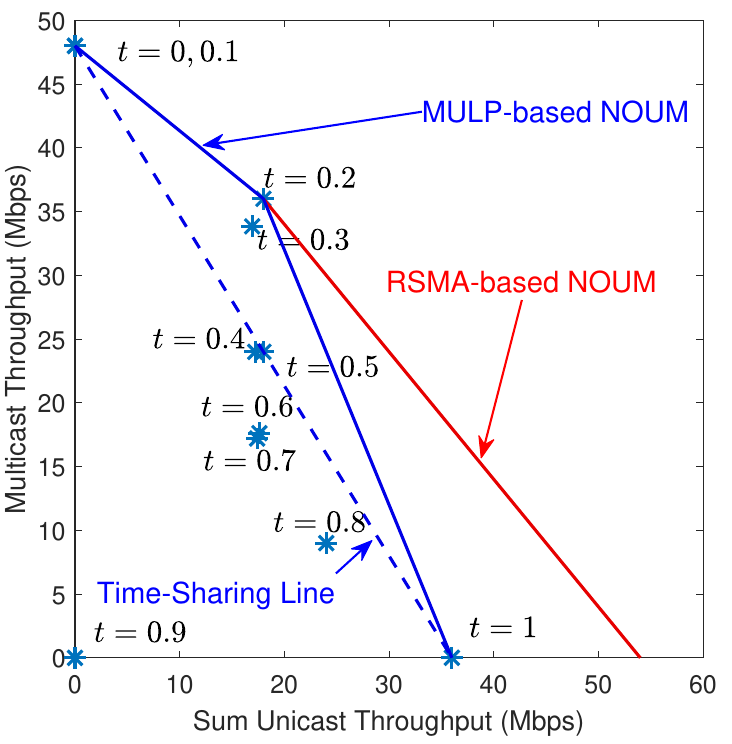}
        \caption{Case 3 ($t^*=0.2$)}
    \end{subfigure}
        \begin{subfigure}{0.24\textwidth}
        \includegraphics[width=\linewidth]{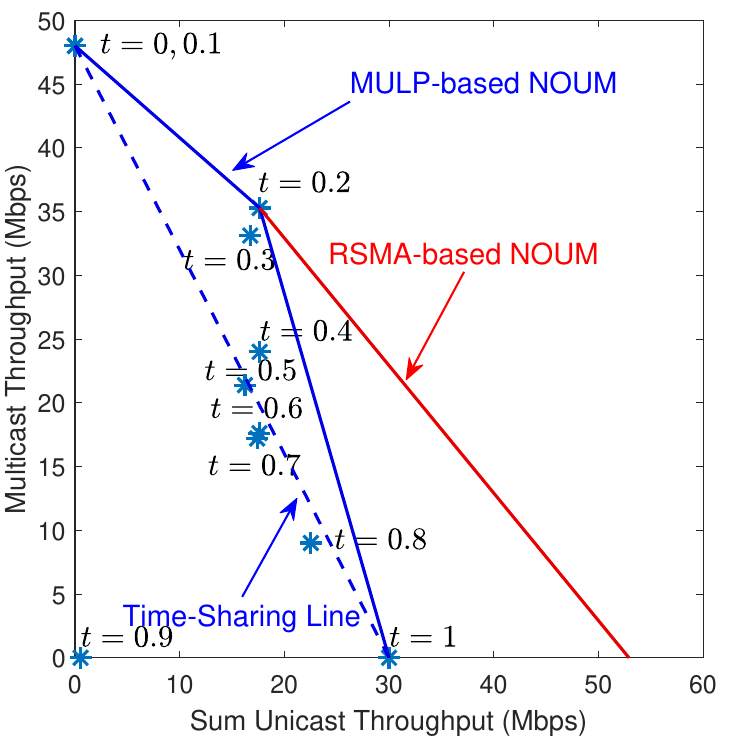}
        \caption{Case 4 ($t^*=0.2$)}
    \end{subfigure}
    \caption{The measured NOUM rate region (i.e., the empirical version of Fig.~\ref{fig:region}). RSMA-based NOUM yields higher unicast throughput than MULP-based NOUM when the multicast data rate is less than 6Mbps (Case 1), 18Mbps (Case 2) and 36Mbps (Case 3, 4).}
    \label{fig: region}
    \end{figure*}
\begin{figure*}[ht]
    \begin{subfigure}{0.24\textwidth}
        \includegraphics[width=\linewidth]{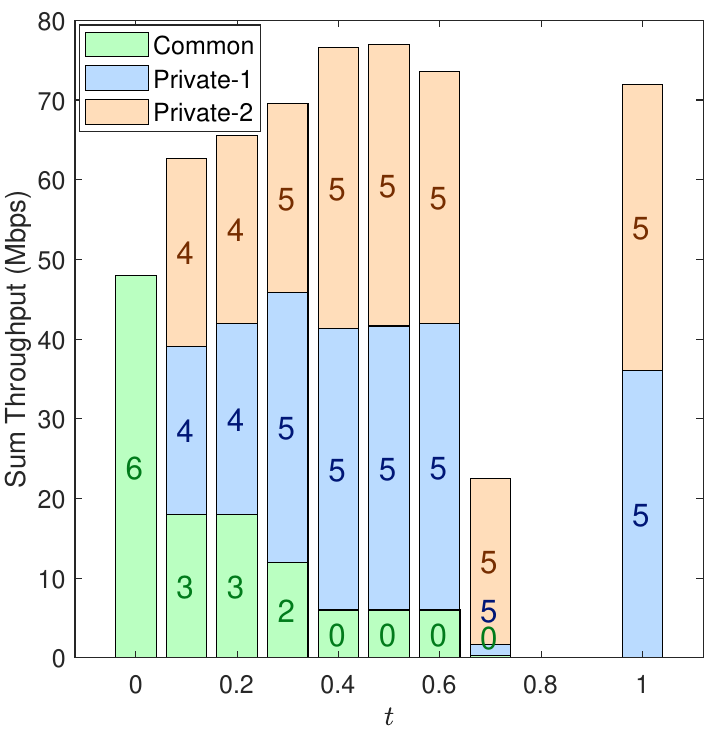}
        \caption{Case 1}
    \end{subfigure}
    \begin{subfigure}{0.24\textwidth}
        \includegraphics[width=\linewidth]{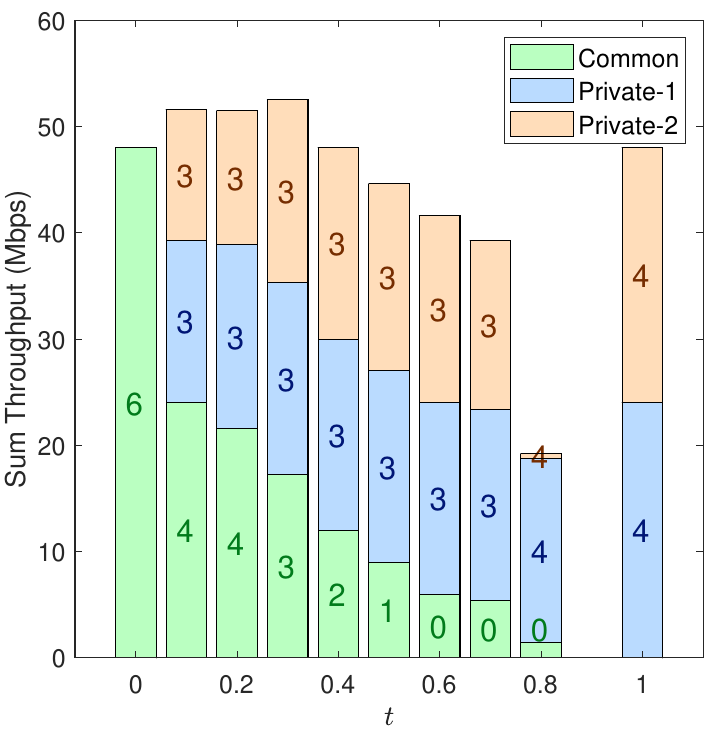}
        \caption{Case 2}
    \end{subfigure}
    \begin{subfigure}{0.24\textwidth}
        \includegraphics[width=\linewidth]{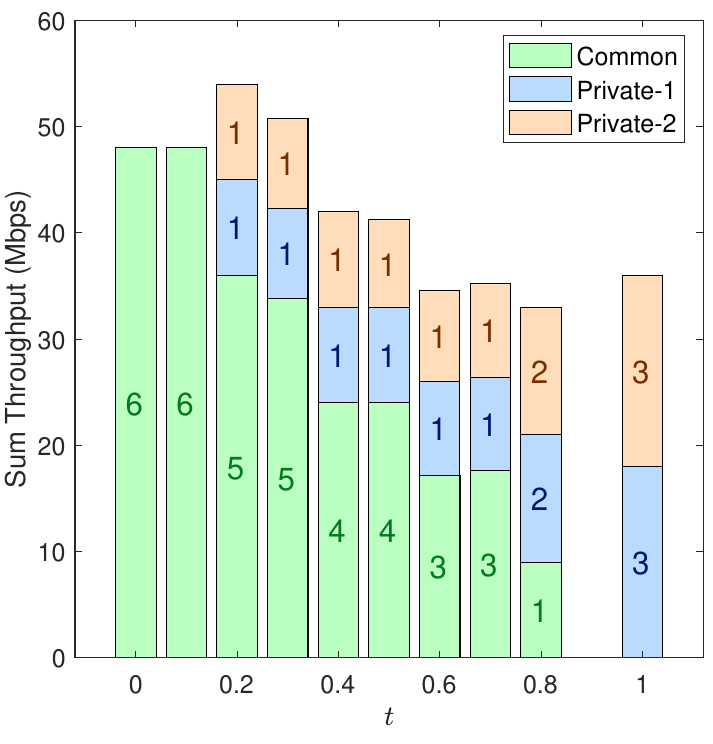}
        \caption{Case 3}
    \end{subfigure}
        \begin{subfigure}{0.24\textwidth}
        \includegraphics[width=\linewidth]{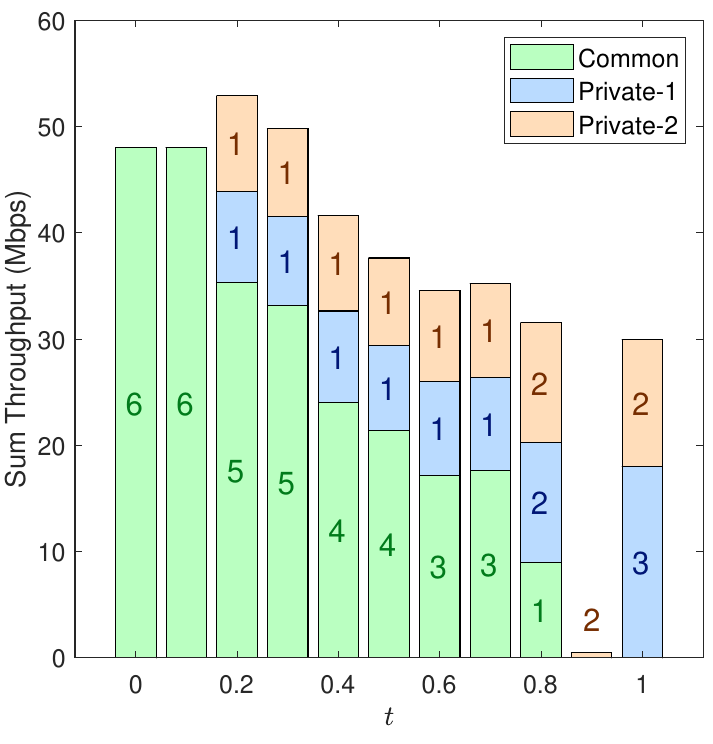}
        \caption{Case 4}
    \end{subfigure}
    \caption{The measured sum throughput broken down in terms of common and private stream contributions. The number within each bar denotes the MCS index from Table~\ref{tab: Mcs table} for the corresponding stream.}
    \label{fig: bar}
\end{figure*}

Fig.~\ref{fig: region} plots the measured NOUM rate region (i.e., the empirical version of Fig.~\ref{fig:region}) for each case, and Fig.~\ref{fig: bar} plots the breakdown of the sum throughput in terms of the common and private stream contributions. We make the following observations: 
\begin{itemize}
    \item[1.] Across Figs.~\ref{fig: region}a-d,  the gap between the rate regions of RSMA- and MULP-based NOUM increases with the channel correlation. As per Lemma \ref{lem1}, RSMA-based NOUM yields higher unicast throughput than MULP-based NOUM when the multicast data rate is less than 6Mbps (case 1), 18Mbps (case 2) and 36Mbps (cases 3, 4).
    \item[2.] Between Fig.~\ref{fig: region}c and d, the main difference occurs at $t=1$, which corresponds to unicast SDMA (Remark \ref{rem:ptB}). The corresponding sum throughput in Case 4 (30Mbps) is lower than in Case 3 (36Mbps) due to the weaker channel of user 2 and the high interference from user 1 due to the high channel correlation. This is evidenced by the lower MCS for user 2 ($\mathcal{I}_2 = 2$ in Fig.~\ref{fig: bar}d v/s $\mathcal{I}_2 = 3$ in Fig.~\ref{fig: bar}c). For $t\neq 1$, the common stream helps the weaker user 2 to partially decode and cancel the high interference it experiences from user 1, leading to improved sum rate performance. Thus, the rate-region for RSMA-based NOUM is identical in Figs.~\ref{fig: region}c and \ref{fig: region}d. This suggests that the channel correlation is a stronger indicator of RSMA-based NOUM's rate performance than the relative pathloss difference between the user channels.
    \item[3.] With increasing channel correlation, the highest sum rate is realized by allocating more power to the common stream. This is evidenced by $t^*$ decreasing from 0.5 (case 1) to 0.3 (case 2) to 0.2 (cases 3, 4). Consequently, the contribution from the common stream to the max. sum throughput increases from 8\% (case 1) to 33\% (case 2) and then to 61\% (cases 3, 4). The more the common stream contributes to the max. sum throughput, the larger the gap between the rate regions of RSMA- and MULP-based NOUM, as there is greater flexibility in using the common stream for multicast and unicast.
    \item[4.] For $t$ approaching but not equal to 1, the measured throughput is 0 (e.g., $t = 0.9$ in all the cases). This is because the low power (only 10\% of the power budget for $t=0.9$) allocated to the common stream is insufficient to decode even the smallest MCS level ($\mathcal{I} = 0$ in Table \ref{tab: Mcs table}). This causes imperfect SIC and error propagation while decoding the private streams.
   \item[5.] For several values of $t$ across Figs.~\ref{fig: region}a-d, the measured multicast and unicast throughputs are below the time-sharing line. This is due to the limited granularity of the MCS levels. As an example, consider the point corresponding to $t = 0.5$ for Case 2 in Fig.~\ref{fig: region}b. One way to push this point above the time-sharing line is by increasing the multicast throughput using a higher MCS level for the common stream. However, from Fig.~\ref{fig: bar}b, the next highest MCS level for the common stream ($\mathcal{I}_C = 2$) is too aggressive, causing imperfect SIC and error propagation. More broadly, points 4 and 5 underscore the importance of the dual assumptions of perfect SIC and MCS levels of arbitrary granularity in Remark~\ref{rem:ptC}.
\end{itemize}

\section{Summary}
In this paper, we characterized and experimentally investigated RSMA-based NOUM for the two-user case. We saw that the benefits of RSMA-based NOUM are more pronounced with increasing channel correlation. This is because the common stream contributes an increasingly larger share to the sum throughput, which provides greater flexibility in using it for multicast and unicast. As a result, the gap between the rate regions of RSMA- and MULP-based NOUM also increases with the channel correlation.

\section{Acknowledgment}
This work was partially funded by UKRI Impact Acceleration Account (IAA) grant EP/X52556X/1 and UKRI grants EP/X040569/1, EP/Y037197/1, EP/X04047X/1, EP/Y037243/1.

\bibliographystyle{IEEEtran}

\end{document}